\documentclass[a4paper,twoside]{article}

\usepackage{epsfig}
\usepackage{subfigure}
\usepackage{calc}
\usepackage{amssymb}
\usepackage{amstext}
\usepackage{amsmath}
\usepackage{amsthm}
\usepackage{multicol}
\usepackage{pslatex}
\usepackage{apalike}
\usepackage{SCITEPRESS} 
\usepackage{pifont} 
\usepackage{stmaryrd} 
\usepackage[small]{caption}

\let\bibhang\relax
\usepackage{natbib} 

\begin{document}

\pubnote{
Preprint  accepted to the 10th Int. Conf. on ENASE. 
Copyright (c) 2015 SCITEPRESS}

\title{Requirements Engineering Aspects\\ of a Geographically Distributed Architecture
}
 
\author{ 
\authorname{Maria Spichkova and Heinz Schmidt}  
\affiliation{RMIT University, Melbourne, Australia}
\email{\{maria.spichkova,  heinz.schmidt\}@rmit.edu.au}
}


\abstract{We present our ongoing work on requirements specification and analysis for the geographically distributed software and systems.  
Developing software and systems within/for different countries or states or even within/for different organisations 
means that the requirements to them can differ in each particular case. 
These aspects naturally impact on the software architecture and on the development process as a whole.
The challenge is to deal with this diversity in a systematic way, avoiding contradictions and non-compliance. 
In this paper, we present a formal framework for the analysis of the requirements diversity, which comes from the differences in the regulations, laws and cultural aspects  for different countries or organisations. The framework also provides the corresponding architectural view and the methods for requirements structuring and optimisation. 
}

\onecolumn \maketitle \normalsize \vfill

\section{\uppercase{Introduction}}
\label{sec:introduction}

Globalisation of the software and systems development  
 offer great  opportunities for the  development industry. 
However, they also mean new challenges coming from the diversity of requirements on different locations, especially in the sense of legal requirements and regulatory compliance.
The reasons for the differences in software and systems requirements within different countries, states and organisations are the cultural and economics diversity, as well as the diversity in standards, legal regulations and laws. 
These challenges have some similarities with the problems of product customisation and the development of product lines, 
but the core of the architecture requirements  is different due the specific nature of the relations between the requirements for the geographically and organisationally distributed development and application of software.

Requirements engineering (RE), i.e., requirements elicitation, evaluation,  specification, and design producing the functional and non-functional requirements, is one of the key disciplines in the software development domain. 
It has a critical impact on the product's quality.  
Requirements-related errors are often a  major cause of the delays in the product delivery and  development costs overruns, cf. e.g., \citep{vanLamsweerde:2008,Pretschner:2007,Sawyer2007}. 
There are several methodologies on development of
software systems from requirements, also enclosing  CASE tools support, e.g., \citep{BroySlotosch2001,holzl2010autofocus,spichkova2011architecture,VerisoftXT_FMDS}.
The RE task is challenging even in the case of a local (non-distributed) development of a product for application within/for a single country or organisation, i.e. 
where the system acts within an environment with a uniform set of standards, legal regulations, etc.  
Thus, in the case of global development we need to have an approach that deals the corresponding issues in a systematic and scalable way. 
There are several approaches that check requirements for compliance \citep{Breaux2008,Relaw2009,Siena2009} or ensuring the compliance of the outcomes of business
processes against outcome-focused regulations \citep{eros2013}.  
We are going a step further, aiming to cover the RE aspects
for the geographically distributed development and application. 
To minimise the overall effort while specifying and also ensuring required software and system requirements in a global development  context,
we  elaborated a logical framework.
The framework provides methodological structuring the requirements for the geographically distributed product development and application, as well as 
developing of the corresponding global architecture requirements.
The purposed approach will help to analyse the relations between requirements    
and to trace requirements' changes  in a global context.

\textbf{Outline:} 
The rest of the paper is structured as follows: 
Section~\ref{sec:related} overviews related work on integration architecture and RE as well as on regulatory compliance within the field of software engineering. 
In Section~\ref{sec:framework} we present our view on the architectural dependencies for the requirements 
 for a global development and geographically distributed product application.
In Section \ref{sec:structure} we introduce the core ideas of requirements structuring process and requirements analysis in a global context. 
Section~\ref{sec:conclusion} concludes the paper by highlighting the main contributions of the paper and describing the future work directions.

\section{\uppercase{Related Work}}
\label{sec:related}

\noindent 
\cite{GlinzRE2007} surveyed the existing definitions of non-functional requirements (NFR), highlights and discusses the problems with the
current definitions, and contributes concepts for overcoming
these problems. 
In our work, we are mainly focusing on NFR  in the sense of legal aspects and regulatory compliance, also taking into account human factor aspects of the requirement modelling~\citep{hffm_spichkova}. 
\\ 

\noindent 
\textbf{Integrating Architecture and Requirements Engineering:} 
The main purpose of the requirements specification (RS) is to elicit and to document the given problem (product/software/system)
using concepts from the problem domain, i.e. on the RE phase we are speaking only on the \emph{problem statement}. 
In contrast to this, the aim of a software  architecture (SA) is to design a \emph{draft of  the solution} for the problem described in the RS, at a high level of abstraction.
Thus, there are  tight interdependencies between
functional/non-functional requirements and architectural elements, which makes  the integration of the RE and architecture 
crucial~\citep{WinWinQR2001,Egyed01refinementand}.
The results of the empirical study conducted by \cite{Ferrari2007} also have shown that  the software architects 
with the knowledge and experience on RE perform better, in terms of architectural
quality, than those without these knowledge end experience.

\cite{Nuseibeh:2001} described a spiral model-like development cycle of requirements and
architecture. 
\cite{Pohl:2007} went further and have provided methodical guidance for the co-design.
An experience-based approach for  integration architecture and RE is presented by \cite{icsePaechKDBKKTPD03}.
This approach that supports
the elicitation, specification and design activity by
providing \emph{experience} in terms of questionnaires,
checklists, architectural patterns and rationale that have
been collected in earlier successful projects and that are
presented to developers to support them in their task.

The REMseS approach \citep{Braun2014}  aims at supporting RE
processes for software-intensive embedded systems. The authors introduced  
fundamental principles of the approach and gave a structural
overview over the guide and the  
tool support.

In contrast to these approaches, our research covers the regulatory compliance aspects and is oriented on legal requirements representation and analysis 
in the scope of the global software development. 
\\

\noindent 
\textbf{Regulatory compliance:}
A survey of efforts to support the analysis of
legal texts  in the context of software engineering is presented in \citep{OttoLegalReq}. 
The authors discuss  the role of law in requirements and 
identify several key elements for any system to support the analysis of
regulatory texts for requirements specification, system
design, and compliance monitoring. 
 \citet{relaw2011} also identified key artefacts,
relationships and challenges in the compliance demonstration of the systemÕs requirements
against  engineering standards and government regulations, also providing
 This work provides a basis for developing compliance meta-model.

\cite{Kiyavitskaya2008}  investigated the
problem of designing regulation-compliant systems and, in
particular, the challenges in eliciting and managing legal
requirements. 
\cite{Breaux2006} reported on an industry
case study in which product requirements were specified
to comply with the U.S. federal laws. 
\cite{Relaw2009} performed a case study using our approach to evaluate
the iTrust Medical Records System requirements for
compliance with the U.S. Health Insurance Portability
and Accountability Act.
\cite{Siena2009LawReq}  presented the guiding rules and a
framework for for deriving compliant-by-construction requirements, also focusing on the U.S. federal laws.

In contrast to these approaches, our research is oriented on global software development, 
dealing with diversity in standards, legal regulations, etc.  within different countries and organisations.


\section{\uppercase{Geographically Distributed Architecture}}
\label{sec:framework}

Suppose that we have to develop $M$ products (software components/systems) $P_1, \dots, P_M$. We denote the set of products by $P$.
Each product $P_{i}$, $1 \le i \le M$ has the corresponding set of requirements $R_{P_{i}}$. 
However, in the case the legal requirements and the regulatory compliance are taken into account, 
the set of requirements will depend on the regulations and laws of the country/state  the product is developed for. 

In the case of global and remote development~\citep{spichkova2013abstract},  we have to deal with cultural and economics diversity, 
which also has an influence on the software and system requirements. 
Suppose the products $P_1, \dots, P_M$ are developed for application in $N$ countries $C_{1}, \dots, C_{N}$ 
with the corresponding
\begin{itemize}
\item  regulations/laws $RegulC_{1}, \dots, RegulC_{N}$  and
\item  
cultural/economics, i.e., human factor  \citep{Borchers:2003},  influences $HFC_{1}, \dots, HFC_{N}$.
\end{itemize}
 
~\\
We denote the set of requirements to the product ${P_i}$ valid for the country $C_{j}$ by $R^{C_{j}}_{P_i}$.
The complete set of requirements to the product ${P_i}$ is then defined by

\begin{equation}\label{LRPi}
R^{P_i}  = \bigcup_{j=1}^{N} R^{C_j}_{P_i}  
\end{equation}
The sets of requirements $R_{P_{i}}$ might be different for each product $P_{i}$ in different countries, i.e. 
$R^{C_{j1}}_{P_i}$ is not necessary equal to $R^{C_{j2}}_{P_i}$ for the case $j1 \neq j2$. 

\begin{figure}[!h] 
  \centering
   {\epsfig{file = 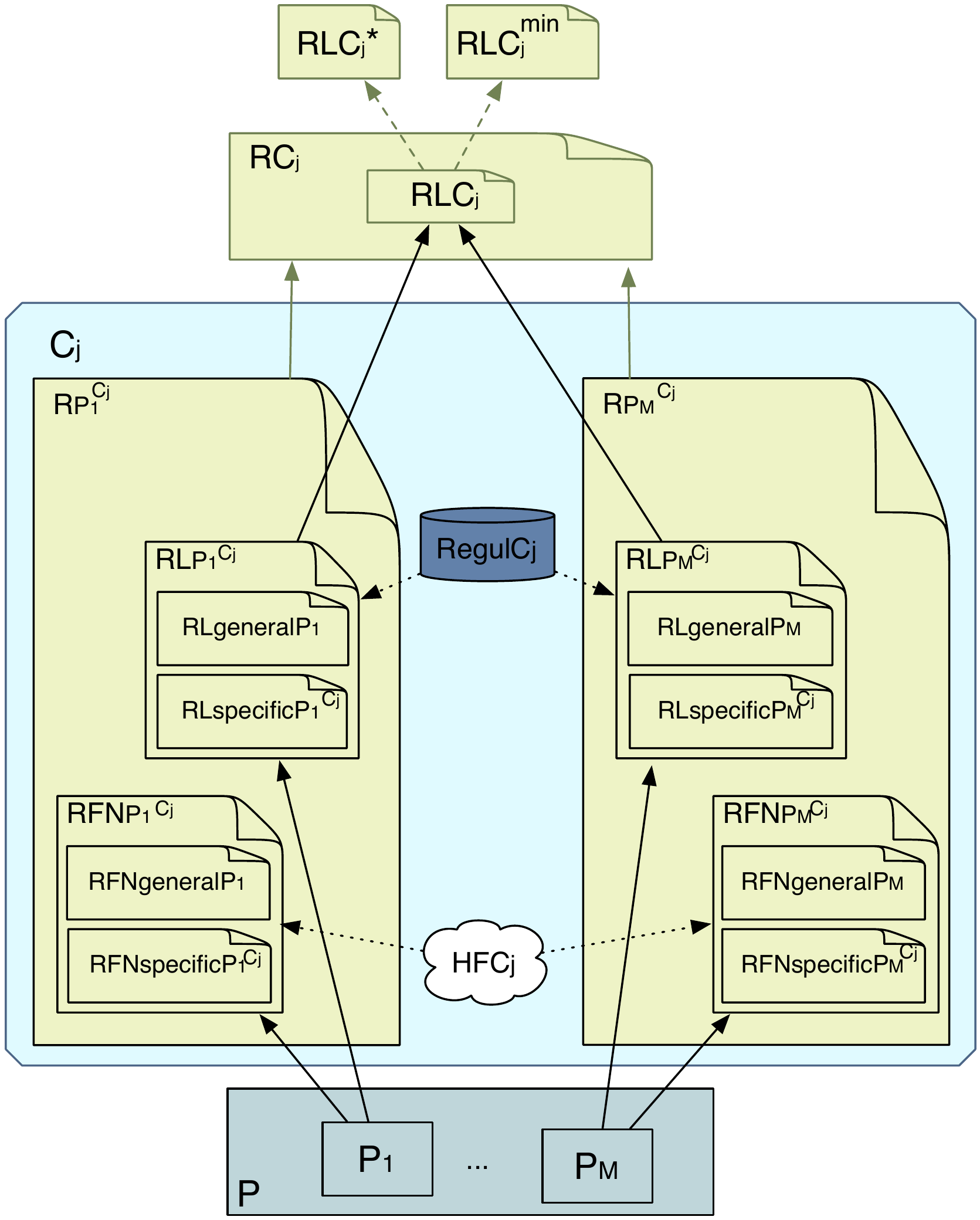, width = 7.5cm}}
  \caption{Architectural dependencies for the requirements based on the regulations and laws of the  country $C_{j}$}
  \label{fig:LRset_Cj}
 \end{figure}

Figure   \ref{fig:LRset_Cj} presents the corresponding architectural dependencies for the requirements based on the regulations and laws of the  country $C_{j}$.
We divide the set of requirements $R_{P_{i}}$  in two (disjoint) subsets. 
For each of these subsets, 
we have to distinguish two separate parts: \emph{general} and \emph{country-specific}:
\begin{itemize} 
\item 
$RL^{C_{j}}_{P_i}$ denotes the requirements based or depending  on the regulations and laws, which could be country/state/organisation-specific. 
Requirements of this kind does not depend on the human factor related aspects. 
\begin{equation}\label{e1}
RL^{C_{j}}_{P_i} =
 RLgeneral^{C_{j}}_{P_i} \cup\ RLspecific^{C_{j}}_{P_i}
\end{equation}
\begin{equation}\label{e2}
RLgeneral^{C_{1}}_{P_i} =   \dots= RLgeneral^{C_{N}}_{P_i}
\end{equation}
\item
 $RFN^{C_{j}}_{P_i}$  denotes the functional and non-functional requirements that are independent from the regulations and laws, but
 may depend on the human factor  related aspects, which could be country-specific.  
\begin{equation}\label{e3}
RFN^{C_{j}}_{P_i} =
 RFNgeneral^{C_{j}}_{P_i} \cup\ RFNspecific^{C_{j}}_{P_i}
\end{equation}
\begin{equation}\label{e4}
RFNgeneral^{C_{1}}_{P_i} =   \dots= RFNgeneral^{C_{N}}_{P_i}
\end{equation}
 \end{itemize}
For simplicity, we denote the general subsets  by $RLgeneral_{P_i}$ and $RFNgeneral_{P_i}$ respectively. 
\\

\noindent
Given the set $RL^{C_j}_{ALL}$ be the set of all legal requirements of the country $C_j$, we can see the set $RL^{C_{j}}_{P_i}$ 
as a \emph{projection} of $RL^{C_j}_{ALL}$ to the product $P_{i}$:
\begin{equation}\label{RLallProj}
RL^{C_{j}}_{P_i}  = RL^{C_j}_{ALL} |_{P_{i}}
\end{equation}

On this basis,  we can say that the set of legal requirements for the country $C_j$ related to the development of products $P_1, \dots, P_M$ is defined as a union over the corresponding requirements sets:
\begin{equation}\label{LRCi}
RL^{C_j}  = \bigcup_{i=1}^{M} RL^{C_j}_{P_i}  
\end{equation}
However, it is not efficient to base the requirements analysis for the geographically distributed development on the sets $RL^{C_j}$:
\begin{itemize}
\item
The sets $RegulC_{1}, \dots, RegulC_{N}$ could have a joint subset $Regul$ of the regulations/laws that are equal for all countries $C_{1}, \dots, C_{N}$:
\begin{equation}\label{Regul}
Regul =  RegulC_{1} \cap\ \dots \cap\ RegulC_{N}
\end{equation}
We denote for each $RegulC_{j}$ the corresponding compliment to $Regul$ (i.e. the country-specific subset) by 
${RegulC_{1}}^{\prime} = RegulC_{1} \setminus Regul$. \\
It it is important to identify these subsets as well as the corresponding subsets of 
$RL^{C_{1}}, \dots, RL^{C_{N}}$, as this helps to trace the $RL$-requirements' changes more efficiently. 
\item
Some requirements in $RL^{C_j}$ can be stronger versions of another requirements from this set. 
For example, if $req_{1} \in RL^{C_{i1}}$ and  $req_{2}  \in RL^{C_{i2}}$ are not equal, they both will belong to $RL^{C_j}$, even if $req_{1}$ is a refinement of $req_{2}$.
In this case, we call $req_{2}$  a \emph{weaker version} of $req_{1}$ and denote it by $req_1 \rightsquigarrow req_2$.
\end{itemize}
Thus we need to have a structured architecture for the legal requirements on the products $P_{1}, \dots, P_{M}$. 
For this reason we build  
the corresponding sets ${RL^{C_j}}^{min}$ and ${RL^{C_j}}^{*}$,  
where 
${RL^{C_j}}^{*}$ denotes the strongest set of  legal requirements for the country $C_j$ (related to the concrete products development),
and $RL_{C_j}^{min}$ denotes the set of legal requirements that should be fulfilled by \emph{all} the products $P_1, \dots, P_M$ developed in the country $C_j$:
\begin{equation}\label{LRCimin}
{RL^{C_j}}^{min}  = RL^{C_j}_{P_1} \cap \dots \cap RL^{C_j}_{P_M}
\end{equation}
${RL^{C_j}}^{*}$ is an optimisation of 
$RL^{C_j}$, where all the weaker versions of the requirements are removed using the algorithm presented in Section \ref{sec:structure}.
%
 
\section{\uppercase{Requirements Structuring and Analysis}}
\label{sec:structure}

In our framework, we perform the analysis based on the optimised views on the requirements sets, focusing on the regulatory/legal aspects. 
First of all, we analyse the sets of relevant regulations  $RegulC_{1}, \dots, RegulC_{N}$. Three cases are possible:
\begin{itemize}
\item
In the case $Regul = \emptyset$, we have the situation when the regulations are completely different  for all $C_{1}, \dots, C_{N}$. 
This also implies that $RLgeneral_{P_i} = \emptyset$ for all $P_{i}$, $1 \le i  \le M$, i.e.,
$RL^{C_{j}}_{P_i} = RLspecific^{C_{j}}_{P_i}$.
We have to trace all the sets $RLspecific^{C_{j}}_{P_i}$ separately:
 changes in $RegulC_{j}$ do not influence on the global development process.
 \item
The regulations are not completely identical  for $C_{1}, \dots, C_{N}$, but $Regul \neq \emptyset$.
 If we can rely on the statical nature of this requirements
  (i.e. that these requirements do not change over the time of the development and the application), 
 it would be beneficial to apply the component-based development paradigm \citep{crnkovic2007component}: 
the requirements $RLgeneral_{P_i}$ or at least the major part of them should correspond to 
 architectural components(s) that are separate from the components corresponding to $RLspecific^{C_{1}}_{P_i}, \dots,  RLspecific^{C_{N}}_{P_i}$.
However, if any of regulations sets 
 $RegulC_{j}$ has some changes, this would influence on the development process as a whole.
 \item
In the case $Regul = RegulC_{1} = \dots = RegulC_{N}$, we have the situation when the regulations are completely identical  for all $C_{1}, \dots, C_{N}$, 
which also means 
 $RL^{C_{j}}_{P_i} = RLgeneral_{P_i}$. 
  If we can rely on the statical nature of this requirements, we have the simplest case for the development process: 
we develop a single component (system) from  $RLgeneral_{P_i}$ to apply if for all $C_{1}, \dots, C_{N}$.
 \end{itemize}

\begin{figure}[!h]
  \centering
   {\epsfig{file = 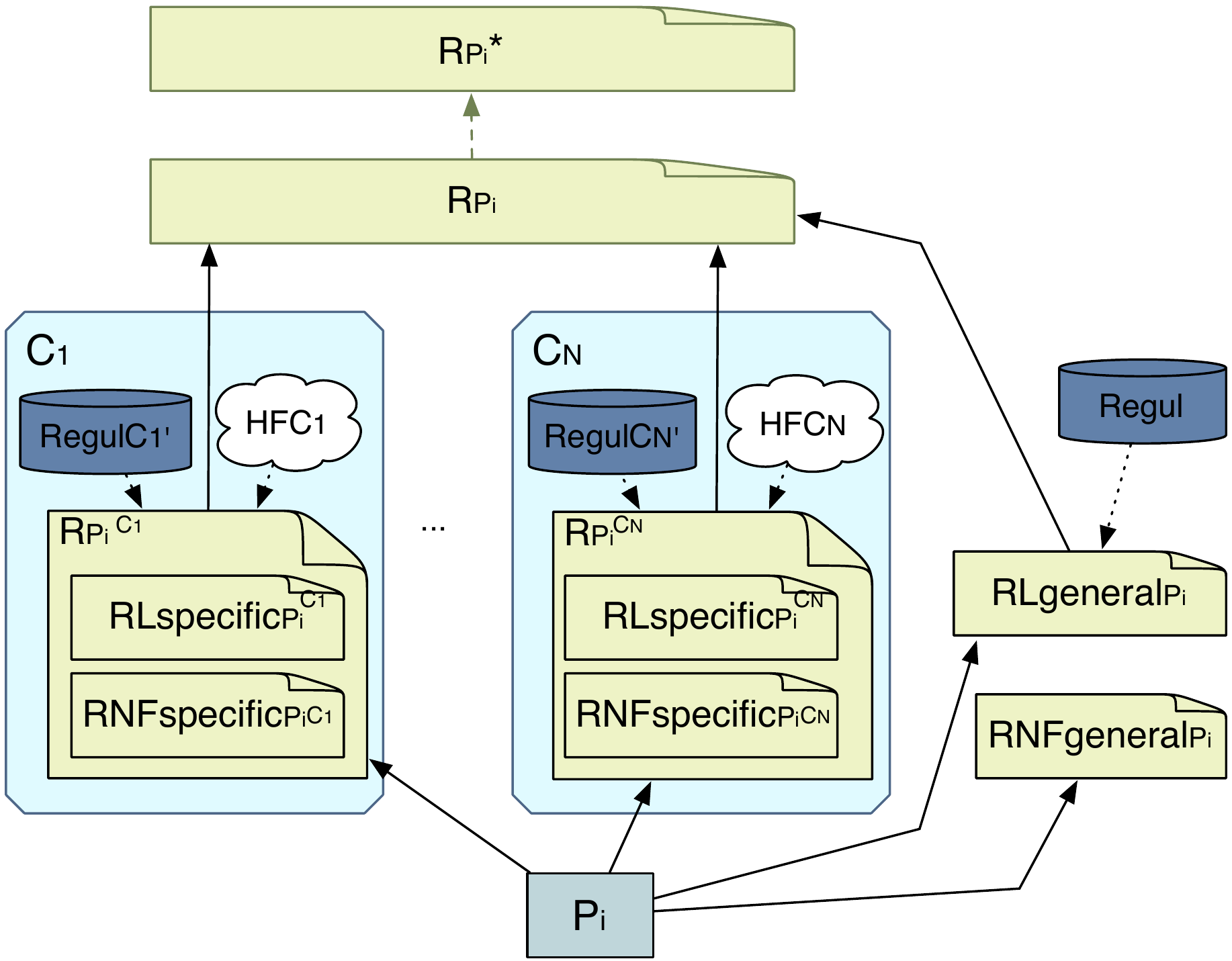, width = 7.5cm}}
  \caption{Architectural dependencies for requirements specified for the product  $P_{i}$}
  \label{fig:LRset_Pi}
 \end{figure}

Thus, ${RL^{C_j}}^{min}$ is defined on basis of $Regul$. 
The corresponding product-centred view on the architectural dependencies is presented on  Figure~\ref{fig:LRset_Pi}. 
The algorithm of constructing ${R^{C_j}}^{min} $ is trivial: we check all the requirements in $R^{C_j}_{P_1}, \dots, R^{C_j}_{P_M}$ 
to find out those elements, which belong to each of the sets.

Our approach is based on the ideas of refinement-based specification and verification.
On the formal level, we need to define which exactly kind of a refinement we mean, e.g., behavioural,
 interface  or conditional refinement 
 \citep{Spichkova2008131,broy_stolen_01}. 
 However, on the level of the logical architecture and modelling of the dependencies between the (sets of) requirements, we can abstract from these details. 

\begin{figure}[!h]
  \centering
   {\epsfig{file = 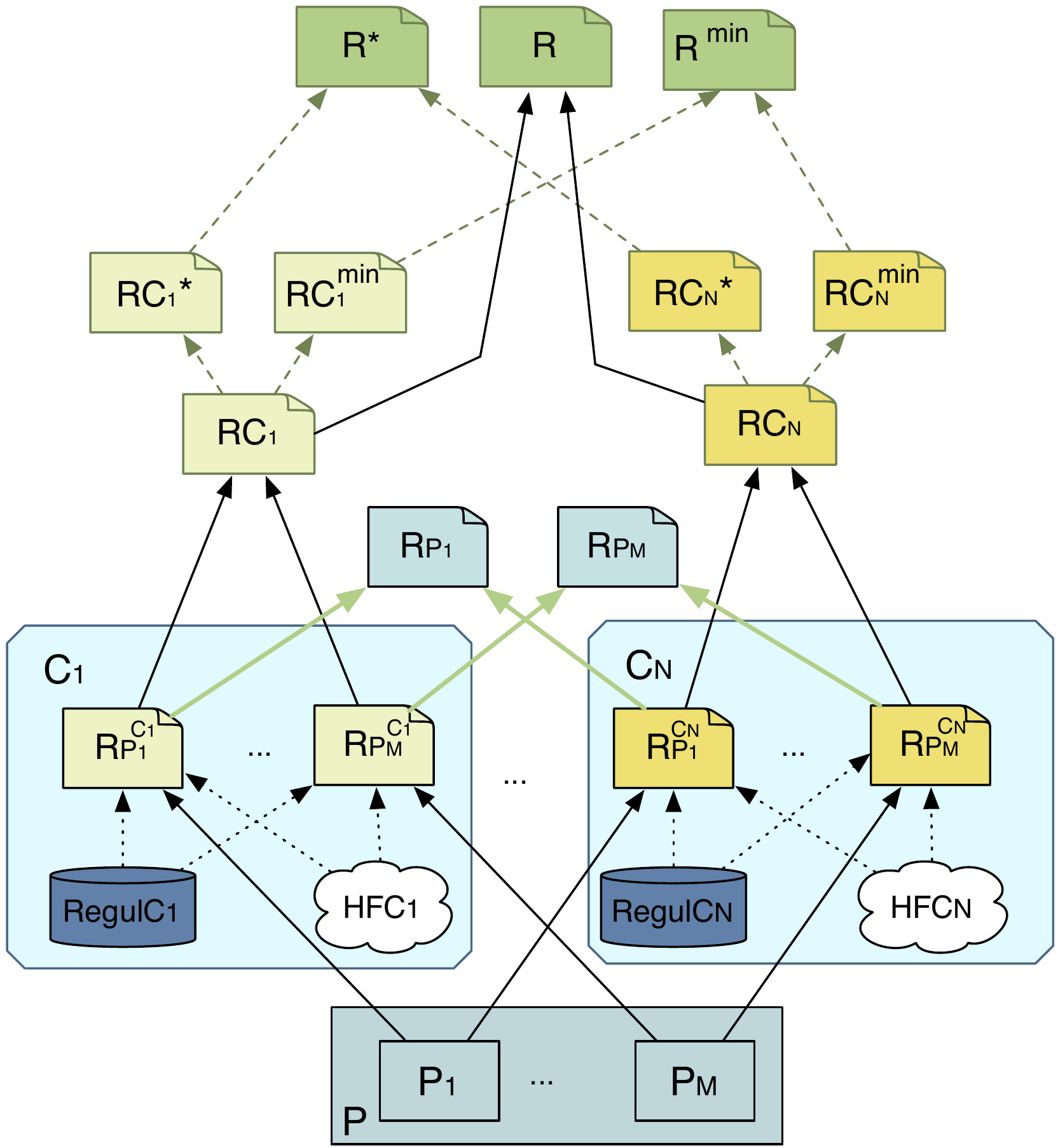, width = 7.5cm}}
  \caption{Requirements structuring for the distributed development of products $P_1, \dots, P_M$}
  \label{fig:LRset}
 \end{figure}

In this paper, we present a simplified version of the optimisation algorithm. 
It can be applied to build the set ${RL^{C_j}}^{*}$ on the basis of $RL^{C_j}$ (cf. Figure~\ref{fig:LRset_Cj} for the country-centred view), 
 to build the set ${R_{P_{i}}}^{*}$ (cf. Figure~\ref{fig:LRset_Pi} for product-centred view) 
as well as to build the strongest global set of requirements $R^{*}$ over 
$C_{1}, \dots, C_{N}$ (cf. Figure \ref{fig:LRset}).  
We start the algorithm with an empty set and build it up iteratively from the elements of the corresponding set. 

\noindent
\emph{Step 0}:  ${RL^{C_j}}^{*} = \emptyset$, $X = {RL^{C_j}}$. 
\\

\noindent
\emph{Step 1}: 
If $X \neq \emptyset$, the ${RL^{C_j}}^{*}$ is complete, otherwise\\
choose a requirement $req \in X$:
\begin{itemize}
\item
If a copy of $req$ already belongs to ${RL^{C_j}}^{*}$ or $r$ is a weaker version of any requirement from ${RL^{C_j}}^{*}$, 
the set  should not be updated on this iteration:\\
$(req \in {RL^{C_j}}^{*}  ~\vee~ \exists y \in {RL^{C_j}}^{*}: req \rightsquigarrow y) \Rightarrow $\\
${RL^{C_j}}^{*} ~ is ~ unchanged$
\item
If $r$ is a stronger version of any requirement(s) from ${RL^{C_j}}^{*}$, 
we add $r$ to ${RL^{C_j}}^{*}$ and remove all the weaker requirements:\\
$\exists y_{1}, \dots, y_{K} \in {RL^{C_j}}^{*}: y_{1}{\rightsquigarrow}req  \wedge\dots \wedge y_{K}{\rightsquigarrow}req) \Rightarrow $
${RL^{C_j}}^{*} =  ({RL^{C_j}}*~ \cup\ req) \setminus \{y_{1}, \dots, y_{k}\}$
\item
If $r$ does not belong to ${RL^{C_j}}^{*}$ and is neither weaker nor stronger version of any requirement from ${RL^{C_j}}^{*}$, 
we add it to ${RL^{C_j}}^{*}$ and proceed the procedure with the next requirement from ${RL^{C_j}}$:\\
$(req \notin {RL^{C_j}}^{*}  ~\wedge$\\
$ \not \exists y \in {RL^{C_j}}^{*}: ( req \rightsquigarrow y \vee y \rightsquigarrow req) \Rightarrow $\\
${RL^{C_j}}^{*} =  ({RL^{C_j}}^{*}~ \cup\ req)$
\end{itemize}

\noindent
\emph{Step 2}: The $req$ element is deleted from the set $X$:\\
$X = X \setminus req$.
\\

\noindent
Steps 1 and 2 are then repeated until $X = \emptyset$.
\\

While identifying $RL^{min}$ we will analyse which products' subcomponents can be build once 
and then reused for the whole product set $P_{1}, \dots, P_{M}$. 
On this basis, we will have more efficient process for the global software and systems development, 
also having an efficient tracing of requirements changes that might come from the changes in the regulations and laws  for  countries $C_{1}, \dots, C_{N}$. 
For example, changes in ${RegulC_{1}}^{\prime}$ imply changes in $RLspecific_{1}$ only, which means that only the country-specific part of the architectural components for $C_{1}$ is affected, where any changes in $Regul$ might influence the global architecture.

While identifying $RL*$ we will obtain the global view on the the products' requirements, which is not overloaded with the variants of the similar requirements,
where some requirements are just weaker versions of other.
%

\section{\uppercase{Conclusions}}
\label{sec:conclusion}

\noindent 
This paper introduces our ongoing work on requirements specification and analysis for the geographically distributed software and systems.  
Developing software and systems within/for different countries or states or even within/for different organisations 
means that the requirements to them can differ in each particular case, which   
naturally impacts on the software architecture and on the development process as a whole.
Dealing with this diversity and avoiding contradictions and non-compliance, is a very challenging and complicated task. A systematic approach is required. 
For this reason, we created a formal framework for the analysis of the software requirements diversity, which comes from the differences in the regulations for different countries or organisations. 
In this paper, we 
$(i)$ presented our architectural dependency model for the requirements on the distributed development and application,
$(ii)$ introduced the core ideas of the corresponding requirements structuring process and requirements analysis in a global context.
$(iii)$ discussed the the research and industrial challenges in this field, as well as discussed our solutions and how they are related to the existing approaches.  

\textbf{Future Work:} 
In our future work we will investigate how to extend the presented ideas to the software and systems development that involves 
 hierarchical dependencies between the sets of regulations/laws.
This could be the case if see the set $C_1, \dots, C_N$ not only as the set of countries/states, but also as a set of organisations having different internal regulations. 
Then we have to deal with hierarchical dependencies with many levels, e.g., 
$(1)$ organisational regulations, $(2)$ state's regulations and laws, $(3)$ country's regulations and laws, where we also need to check which of the regulations are applicable in each particular case.

\vfill
\bibliographystyle{apalike}
{\small

}

\vfill
\end{document}